\documentclass[aps,superscriptaddress,reprint,floatfix]{revtex4-1}

\usepackage[plainpages=false, colorlinks=true, anchorcolor=blue, linkcolor=blue, citecolor=blue, bookmarks=false]{hyperref}
\usepackage{amsmath, amsthm, amssymb}
\usepackage[utf8]{inputenc}
\usepackage{empheq}
\usepackage{comment}
\usepackage{graphicx}
\usepackage{hyperref}
\usepackage{siunitx}
\usepackage{xcolor} %Use to highlight comments in colour
\usepackage{dblfloatfix}

\newcommand{\gagg}{g_{a\gamma\gamma}} %_{a\gamma\gamma}

\begin{document}
\title{Parametric co-linear axion photon instability}

\author{K. A. Beyer}
\email[Authors to whom correspondence should be addressed: ]{konstantin.beyer@physics.ox.ac.uk}
\affiliation{Department of Physics, University of Oxford, Parks Road, Oxford OX1 3PU, UK}
\author{G. Marocco}
\email{giacomo.marocco@physics.ox.ac.uk}
\affiliation{Department of Physics, University of Oxford, Parks Road, Oxford OX1 3PU, UK}
\author{C. Danson}
\affiliation{AWE, Aldermaston, Reading RG7 4PR, UK}
\affiliation{OxCHEDS, Clarendon Laboratory, Department of Physics, University of Oxford, UK}
\author{R. Bingham}
\affiliation{Rutherford Appleton Laboratory, Chilton, Didcot OX11 0QX, UK}
\affiliation{Department of Physics, University of Strathclyde, Glasgow G4 0NG, UK}
\author{G. Gregori}
\affiliation{Department of Physics, University of Oxford, Parks Road, Oxford OX1 3PU, UK}
\begin{abstract}
   Axions and axion-like particles generically couple to QED via the axion-photon-photon interaction. This leads to a modification of Maxwell's equations known in the literature as axion-electrodynamics. The new form of Maxwell's equations gives rise to a new parametric instability in which a strong pump decays into a scattered light wave and an axion. This axion mode grows exponentially in time and leads to a change in the polarisation of the initial laser beam, therefore providing a signal for detection. Currently operating laser systems can put bounds on the axion parameter space, however longer pulselengths are necessary to reach the current best laboratory bounds of light-shining through wall experiments.
\end{abstract}

\maketitle
%-----------------------------------------
\section{Introduction}
Despite the many successes of the Standard Model, it is thought to be incomplete. The strong sector of the Standard Model allows CP violation, which is constrained to cancel the electroweak contribution by measurements of the electromagnetic dipole moment of the neutron \cite{Baker:2006ts,Afach:2015sja}. This is known as the strong CP problem, which can be solved in an elegant fashion by introducing a new chiral, anomalous $U(1)_\text{PQ}$ symmetry, as pointed out by Peccei and Quinn \cite{PhysRevLett.38.1440,PhysRevD.16.1791}. Such global symmetry, after spontaneously breaking at a high scale, dynamically drives the CP violating parameter $\Bar{\theta}$ to $0$. The pseudo Nambu-Goldstone boson originating from the symmetry breaking process is called the axion \cite{Weinberg:1977ma,Wilczek:1977pj}. It was pointed out that axion models can account for the dark matter content of the universe \cite{Preskill:1982cy,Abbott:1982af,Dine:1982ah}. Axion-like particles (ALPs) may abundantly arise in extensions of the Standard Model, such as the low energy spectrum of String Theory \cite{Witten:1984dg,Arvanitaki:2009fg}. In the following we will use the term axion to refer to both, the CP restoring QCD axion and ALPs.

Axions generically couple to Standard Model photons via a dimension 5 operator 
\begin{equation}
    \label{Eq:AxEMVertex}
    \mathcal{L}_{a\gamma\gamma}=\gagg a\mathbf{E}\cdot\mathbf{B}.
\end{equation}
with a coupling strength parametrized by $\gagg$.  The inclusion of equation \ref{Eq:AxEMVertex} modifies Maxwell's equations \cite{Wilczek:1987mv}. Axion electrodynamics in vacuum has been widely applied to axion detection experiments (for a review see, e.g., \cite{Sikivie:2020zpn}). A range of phenomena like the axion-photon mass mixing in magnetic field backgrounds \cite{Sikivie:1983ip,Raffelt:1987im} or axion sourced birefringence \cite{1986PhLB..175..359M} are extensively studied in the literature. 

 %Axion electrodynamics in matter has received much less attention than the in vacuum case. A key difference is the presence of plasmons, which are of importance  due to their resonant mixing with axions \cite{Mikheev:1998bg,Das:2004ee, Ganguly:2008kh, Caputo:2020quz}. The dispersion relations were further studied in \cite{Visinelli:2018zif}. More recent work has focused on exploiting in medium effects to detect any axions comprising the dark matter \cite{TheMADMAXWorkingGroup:2016hpc, Millar:2017eoc, Lawson:2019brd}. Additionally, work has been done using modified version of axion electrodynamics in which electric-magnetic duality is imposed \cite{VISINELLI:2013fia}. Various lab-based axion detection schemes have been proposed based on these equations \cite{Tercas:2018gxv,Mendonca:2019eke}. 

%Many axion-plasmon interactions remain so far ignored. The interaction of intense laser beams with a plasma is known to generate many instabilities. Stimulated Raman scattering (SRS) is an example where an intense laser beam resonantly scatters from electron density fluctuations into a scattered  electromagnetic wave and a plasmon. The beating of the incident beam with the scattered beam produces a ponderomotive force enhancing the density fluctuation, creating a positive feedback loop \cite{,1972PhFl...15..446K,1974PhFl...17..778D,1975PhFl...18.1002F} (for a complete picture, see chapter 7 of \cite{1988aw.....73.....K} and references therein).

The propagation of an intense laser beam receives corrections due to the presence of axions. As was investigated by the PVLAS collaboration, axions source birefringence \cite{DellaValle:2015xxa} of light propagating through a strong constant magnetic field. This is a consequence of the altered dispersion relation \cite{Raffelt:1987im}. In this work we extend previous analyses to look at systems in plane-wave backgrounds. We will demonstrate the existence of a new parametric decay instability of one photon in the seed pulse into an axion and a secondary photon, satisfying
\begin{equation}
    \label{Eq:ModeMatching}
    \omega_0=\omega_\gamma+\omega_a\qquad\text{and}\qquad\mathbf{k}_0=\mathbf{k}_\gamma+\mathbf{k}_a.
\end{equation}
Here, subscript $\gamma$ denotes the scattered photon and $a$ the axion. Physically, small fluctuations in the axion field source electromagnetic fields. Such fields then interact with the pump pulse and source the axion field. Hence, a positive feedback is generated which leads to the growth of the axion fluctuations. A similar idea was investigated in \cite{Mendonca:2007zz} for the case of parametric excitation of axions, where a different ansatz is used for the time-dependence of the fields. Additionally, the stimulated decay of photons (or more specifically, plasmons) into axions have been considered in \cite{1988PhRvD..37.1356R}.

Decay of a photon into a secondary photon and a massive axion in vacuum is not possible because of energy momentum conservation, however the intense background beam changes the dispersion relation thereby allowing such processes. From a field theoretic point of view, photons in a background of a highly occupied photon beam may decay into massive particles non-perturbatively \cite{Arza:2020zop}.

The polarisation-dependent coupling \eqref{Eq:AxEMVertex} forces the scattered mode to have a polarisation orthogonal to the seed pulse. Hence, the laser polarisation depletes. This rate of polarisation change scales as $\gagg$, as opposed to the $\gagg^2$ scaling of conventional birefringence or dichroism experiments \cite{1986PhLB..175..359M}.

This paper is organised as follows: In section \ref{Sec:AxPlCoupling} we derive the set of equations coupling the axion to electrodynamics in a strong pump background and combine them to find the dispersion relation of the system. In section \ref{Sec:DispersionRel} we solve the dispersion relation. For simplicity we limit our analysis to the co-linear limit and leave the general treatment for a later paper. In section \ref{Sec:Disc} we comment on the significance of this axion-photon instability. Within the paper we employ natural units with $\hbar=c=k_B=4\pi\varepsilon_0=1$.

%-------------------------------------------------------
%-------------------------------------------------------
\section{\label{Sec:AxPlCoupling}Axion-photon dispersion relation}
The axion-photon vertex \eqref{Eq:AxEMVertex} introduces a source and current to the vacuum Maxwell's equations \cite{Sikivie:1983ip,Wilczek:1987mv}
\begin{gather}
\label{Eq:Maxwell}
\begin{aligned}
    \nabla\cdot\mathbf{E}&=\gagg\left(\nabla a\right)\cdot\mathbf{B},\\
    \nabla\cdot\mathbf{B}&=0,\\
    \nabla\times\mathbf{E}&=-\partial_t\mathbf{B},\\
    \nabla\times\mathbf{B}&=\partial_t\mathbf{E}+\gagg\left(\mathbf{E}\times\nabla a - \mathbf{B}\partial_t a\right).
\end{aligned}
\end{gather}
The electric and magnetic fields in the axion photon coupling term (\ref{Eq:AxEMVertex}) are defined as
\begin{gather}
    \label{Eq:PotentialDef}
\begin{aligned}
    \mathbf{E}&=-\partial_t\mathbf{A}-\nabla\phi,\\
    \mathbf{B}&=\nabla\times\mathbf{A}.
\end{aligned}
\end{gather}
We employ Coulomb gauge $\nabla\cdot\mathbf{A}=0$, and work in the \emph{colinear} limit where all momenta, are parallel to find the wave equations for the gauge potentials
\begin{equation}
    \label{Eq:EOM-A}
    \left(\partial_t^2 - \nabla^2 \right)\mathbf{A}=\gagg\nabla\times \left(a\partial_t\mathbf{A}\right)
\end{equation}
and
\begin{equation}
    \label{Eq:EOM-phi}
    \nabla^2\phi=0.
\end{equation}
In this colinear limit, we may use the residual gauge freedom to set $\phi = 0$. The remaining equation of motion for the axion is then\begin{equation}
    \label{Eq:EOM-a}
    \left(\partial_t^2-\mathbf{\nabla}^2+m_a^2\right)a=\gagg(\partial_t\mathbf{A})\cdot(\nabla\times\mathbf{A}).
\end{equation}

We perform a linear stability analysis by setting
\begin{equation}
    \begin{split}
    \mathbf{A} = \mathbf{A}_0 \cos\left(\omega_0t-\mathbf{k}_0\cdot\mathbf{x}\right) + \tilde{\mathbf{A}},\qquad
    a = \tilde{a},
    \end{split}
\end{equation}where $\Tilde{\mathbf{A}}, \,\tilde{a} \ll\mathbf{A}_0$. Note that this assumption is only valid for early times before the instability turns strongly non-linear. To solve this system of coupled partial differential equations, we first perform a spatial Fourier transform of all fields
\begin{equation}
    \label{Eq:SpatFour}
    f\left(\mathbf{k},t\right)\equiv\int f\left(\mathbf{x},t\right)e^{-i\mathbf{k}\cdot\mathbf{x}}\frac{d^3\mathbf{x}}{\left(2\pi\right)^{3/2}}
\end{equation}
to find a coupled set of linearised ordinary differential equations
\begin{widetext}
\begin{align}
    \label{Eq:Four-A}
    \left(-\partial_t^2-\mathbf{k}^2\right)\Tilde{\mathbf{A}}\left(\mathbf{k},t\right)&=-\gagg\frac{\omega_0}{2}\left(\mathbf{k}\times\mathbf{A}_0\right)\left(\Tilde{a}_- e^{-i\omega_0t} - \Tilde{a}_+ e^{i\omega_0t}\right),
    \end{align}
    \begin{align}
    \label{Eq:Four-a}
    \left(-\partial_t^2-\mathbf{k}^2-m_a^2\right)\Tilde{a}\left(\mathbf{k},t\right)=&\frac{\gagg}{2}\mathbf{A}_0\cdot\left[e^{i\omega_0t}\left(\omega_0\mathbf{k}_+-i\mathbf{k}_0\partial_t\right)\times\Tilde{\mathbf{A}}_+-e^{-i\omega_0t}\left(\omega_0\mathbf{k}_--i\mathbf{k}_0\partial_t\right)\times\Tilde{\mathbf{A}}_-\right].
    \end{align}
\end{widetext}
with subscript $\pm$ denoting the dependency on $\mathbf{k}\pm \mathbf{k}_0$.

To analyse the coupled system's behaviour we make an Ansatz for the temporal dependency of the fields
\begin{equation}
    \label{Eq:FieldAnsatz}
    f\left(\mathbf{k},t\right)=f\left(\mathbf{k}\right)e^{-i\omega\left(\mathbf{k}\right)t}
\end{equation}
which reduces the differential equations (\ref{Eq:Four-A}) and (\ref{Eq:Four-a}) to algebraic equations in the fields. They are solved for frequencies satisfying the dispersion relation
\begin{equation}
    \label{Eq:AxColinear}
    D^a(\omega, k)=\frac{\gagg^2A_0^2}{4}\omega_0\left(\omega_0 k-\omega(k) k_0\right)\left(\frac{k_-}{D^\gamma_-}+\frac{k_+}{D^\gamma_+}\right),
\end{equation}
where $k=|\mathbf{k}|$, $D^a(\omega, k)\equiv \omega^2-k^2-m_a^2$, $D^\gamma_\pm\equiv(\omega\pm\omega_0)^2-(k\pm k_0)^2$, and functions of $(\omega\pm2\omega_0,\mathbf{k}\pm 2\mathbf{k}_0)$ are neglected as off-resonant and hence sub-dominant.

The rest of the paper is concerned with an investigation of this dispersion relation.

%--------------------------------------------------------
%--------------------------------------------------------
\section{\label{Sec:DispersionRel}Growth Rate}
We would now like to solve equation \eqref{Eq:AxColinear} for real $k$, while allowing complex $\omega$. The dispersion relation is a $6^\text{th}$-order polynomial in $\omega$. Immediately the two real roots $\omega(k)=k$ can be discarded in the following as we are interested in unstable modes. Such unstable modes are characterised by a complex frequency with non-vanishing imaginary part $\Gamma\equiv\Im\left(\omega\right)$ which defines the growth rate. We proceed by working with the remaining $4^\text{th}$-order polynomial and splitting $\omega$ into its real and imaginary parts $\omega(k)=\omega_a(k)+i\Gamma(k)$ for $\omega_a(k),\Gamma\in\mathcal{R}$. The resulting real and imaginary equations are linearly independent and must be satisfied individually.

\begin{figure}
    \centering
    \includegraphics[width=0.475\textwidth]{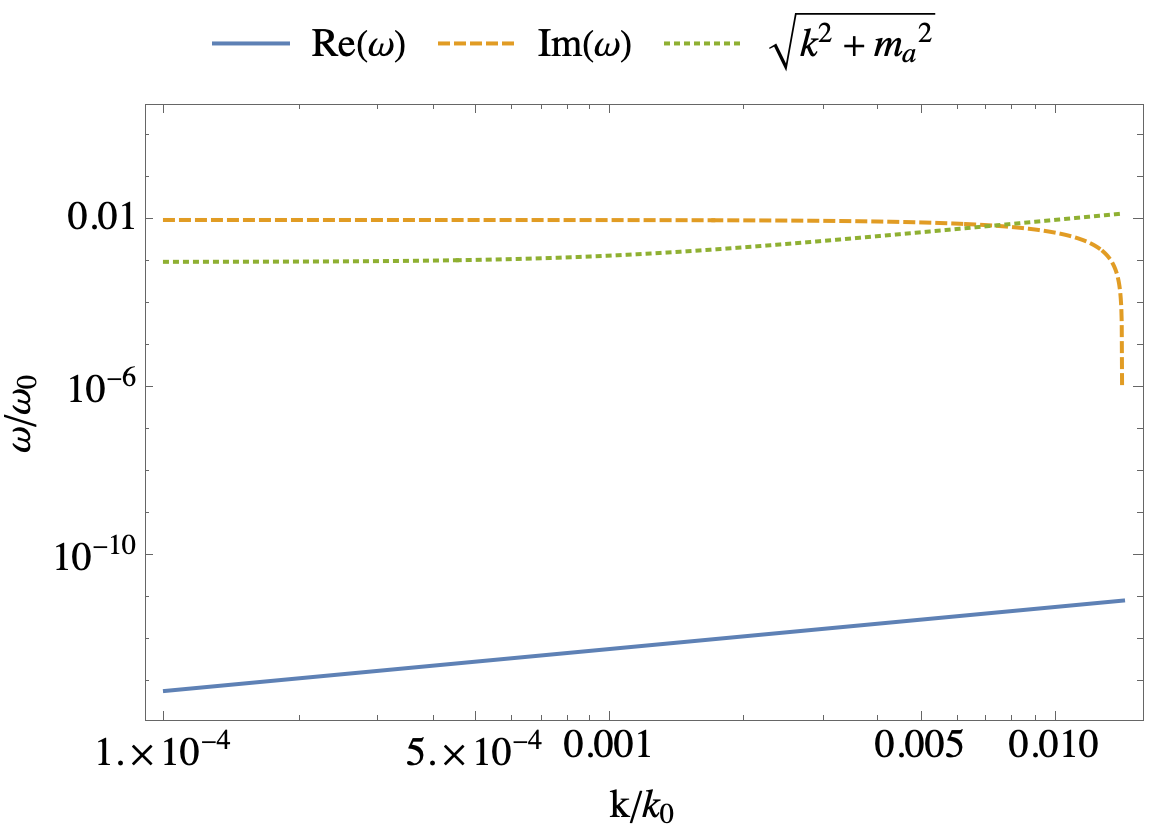}
    \caption{The plot shows the growing solution to the dispersion relation (\ref{Eq:AxColinear}). The blue solid curve shows the real part $\omega_a(k)$ corresponding to the modes frequency while the dashed yellow curve depicts the growth rate $\Gamma$. Note that here we included the $\mathcal{O}(k^2)$ contributions to show the cutoff. As a reference we include the vacuum dispersion relation for an axion with the same mass in dotted green. For illustration purposes we have chosen $m_a/\omega_0=\SI{e-3}{}$ and $\gagg A_0=\SI{e-3}{}$. We see that even for such large values, the axion frequency is negligibly small.}
    \label{fig:DispRelNum}
\end{figure}

The imaginary equation is a cubic polynomial in the growth rate $\Gamma$ with one trivial solution $\Gamma=0$ and two non-trivial ones
\begin{widetext}
\begin{equation}
    \label{Eq:GrowthSol}
    \Gamma=\pm\sqrt{\frac{3 k \omega_a ^2+2 \omega_a ^3-4\omega_a\omega_0^2-k^3-m_a^2 (\omega_a+k)+\gagg^2 A_0^2 k\omega_0^2 }{k-2 \omega_a }}.
\end{equation}
\end{widetext}
From our ansatz \eqref{Eq:FieldAnsatz} we know that a positive $\Gamma$ corresponds to growth, hence we focus only on the positive root. We know that the axion field grows because of the positive feedback through the $\mathbf{E}\cdot\mathbf{B}$ coupling which feeds energy into the axion mode at the expense of the seed beam. After substitution of $\Gamma$ into the real part of the equation we are then left with a single algebraic equation for $\omega_a(k)$. Figure \ref{fig:DispRelNum} shows that in our case, the frequency $\omega_a(k)$ is very small and thus motivating an expansion in $\omega_a$, which we find to be
\begin{equation}
    \label{Eq:WaSolution}
    \omega_a(k)=k\frac{(\gagg A_0)^2}{4} \left(\frac{(\gagg A_0)^2}{4}-\left(\frac{m_a}{\omega_0}\right)^2\right).
\end{equation}
Here we dropped terms of order $6$ in $(\gagg A_0)$, $(m_a/\omega_0)$ and $k$ as being small. We may then express the growth rate as
\begin{equation}
    \label{Eq:GrowthSolFull}
    \Gamma(k)=\omega_0\sqrt{\frac{(\gagg A_0)^2}{4}-\left(\frac{m_a}{\omega_0}\right)^2-\left(\frac{k}{k_0}\right)^2},
\end{equation}
where we have again dropped higher order terms in $(\gagg A_0)$, $(m_a/\omega_0)$ and $(k/k_0)$. 

Upon a closer look at \eqref{Eq:GrowthSolFull} the qualitative behaviour of the numerical solution depicted in fig \ref{fig:DispRelNum} is recovered. We find growth below a cutoff in $k$ which is defined by
\begin{equation}
    \label{Eq:kCutoff}
    k_\text{cutoff}=k_0\sqrt{\frac{(\gagg A_0)^2}{4}-\left(\frac{m_a}{\omega_0}\right)^2}
\end{equation}
at which point $\Gamma$ becomes imaginary and our solution breaks down because the two equations we generate from \eqref{Eq:AxColinear}, one real and one imaginary, are only linearly independent for $\omega_a(k),\Gamma\in\mathcal{R}$. Once the above threshold is passed, the only solution satisfying this criterion has $\Gamma=0$ which matches the numerical solution. The growth rate is essentially constant in $k<k_\text{cutoff}$ and in the following we will suppress the dependence of $\Gamma$. \eqref{Eq:GrowthSolFull} also reveals a second cutoff for $m_a/\omega_0>\gagg A_0$ above which no instability is found.

From \eqref{Eq:WaSolution} we know the phase velocity of the unstable modes
\begin{equation}
    \label{Eq:PhaseVelocity}
    v_a=\frac{\partial\omega_a(k)}{\partial k}=\frac{(\gagg A_0)^2}{4} \left(\frac{(\gagg A_0)^2}{4}-\left(\frac{m_a}{\omega_0}\right)^2\right),
\end{equation}
which can be much less than the speed of light even for massless $m_a=0$ axions.

%--------------------------------------------------
\section{\label{Sec:Disc}Discussion and application}
A possible search experiment based on the above discussion utilises lasers to drive the instability. The frequency of any growing axion mode is very small and thus does not greatly affect the frequency or total energy of the pump beam. The most suitable observable is instead the polarisation of the pump beam. By the nature of the $\mathbf{E}\cdot\mathbf{B}$ coupling, any seed-photon which decays into an axion, will also produce a photon of opposite polarisation.

The set-up in mind consists of a pump pulse polarised in $\hat{x}$ and a weaker probe with polarisation in $\hat{y}$. As long as no axion field is present, $\mathbf{E}\cdot\mathbf{B}=0$ for two colinear electromagnetic plane waves. As soon as the axion field is present, the equations of motion for the electromagnetic fields are no longer linear and $\mathbf{E}\cdot\mathbf{B}\neq 0$. Because realistic laser fields are focused, there will always be modes present with a small but nonzero angle between the momenta. Two such modes have a non-zero $\mathbf{E}\cdot\mathbf{B}$ sourcing an axion field. The part of the field co-linear with the two initial laser beams is then well described by our equations and the axion field grows. 

The axion field is sourced from the coupling of a mode from the seed pulse with frequency $\omega_0$ and one from the probe at $\omega_1=\omega_0+\omega_a$. The axion field's source at the difference frequency is then \eqref{Eq:EOM-a}
\begin{equation}
    \label{Eq:AxFieldInitialSource}
    \gagg\omega_0\mathbf{A}_0 \cdot\left[\left( \mathbf{k_0}+\mathbf{k}_1\right)\times\tilde{\mathbf{A}}\right]e^{i\omega_a t-i(\mathbf{k}_1-\mathbf{k}_0)\cdot\mathbf{x}}
\end{equation}
where we used the fact that $\omega_a\ll\omega_0$ and assumed $\tilde{\mathbf{A}}(\omega_0)=\tilde{\mathbf{A}}(\omega_0+\omega_a)$, an assumption which is justified once again by the smallness of $\omega_a$. For this axion mode to cross the unstable dispersion shown in \ref{fig:DispRelNum}, we require the momentum to fall below the threshold \eqref{Eq:kCutoff} which translates into an upper bound for the collision angle $\alpha$
\begin{equation}
    \label{Eq:AngleReq}
    \left\lvert\mathbf{k}_1-\mathbf{k}_0\right\rvert\simeq\sqrt{2\omega_0^2\left(1-\cos\alpha\right)}\simeq\omega_0\alpha < \frac{\gagg A_0 \omega_0}{2}.
\end{equation}
The upper bound on the collision angle is very small, hence there will exist two photons colliding at such an angle in any realistic focused laser even if the focus is very long. Hence, two photons of appropriate frequency $\omega_0-\omega_1=\omega_a$ colliding at an angle $\alpha <\gagg A_0/2$, source an axion field whose dispersion matches the unstable dispersion \eqref{Eq:WaSolution}. Because $\omega_a\ll\omega_0$ the two photons almost have the same frequency and if they collide symmetrically, the resulting axion is co-linear with the beam axis. From here our above equations apply and describe the growth of this field.

To estimate the change in polarisation we start by defining the initial polarisation of the electromagnetic field produced by the strong pump and weaker probe beam. The superposed gauge field is $\mathbf{A}_0+\tilde{\mathbf{A}}_\pm$ and we define the resulting electric field's polarisation plane to be $\hat{x}$. The probe field grows with the axion field and therefore the polarisation plane changes to $\hat{x}-\varepsilon e^{\Gamma t}\hat{y}$
which is at an angle
\begin{equation}
    \label{Eq:PolAngle}
    \vartheta\sim \varepsilon e^{\Gamma t}
\end{equation}
compared to the initial polarisation. Here, $\varepsilon\propto (\gagg A_0)^3$, one factor of the coupling from the source of equation \eqref{Eq:EOM-A}, one from the axion source \eqref{Eq:EOM-a} and one from the angle $\alpha$ between the seeding photons \eqref{Eq:AngleReq}. The rest of the functional dependence can in principle be obtained from the equations of motion, however as will be clear below, they do not affect the achievable bounds significantly because the most important part is the exponential. The timescale of growth $t$ will be set by the laser pulse length $\tau$. The growth rate \eqref{Eq:GrowthSolFull} depends on the laser parameters
\begin{equation}
    \label{Eq:EfoldGrowth}
    \Gamma=\SI{3e7}{s^{-1}}\left(\frac{\gagg}{\SI{e-8}{GeV^{-1}}}\right)\left(\frac{\mathcal{I}}{\SI{e23}{W/cm^2}}\right)^\frac{1}{2}.
\end{equation}
The intensity achievable depends on the focal spot size and must be maintained for long enough distances to encompass the wavelength of the unstable axion mode $\lambda_a\geq k_\text{cutoff}^{-1}$. We can estimate that distance by
\begin{equation}
\label{Eq:AxWAvelength}
    \lambda_a\geq\SI{10}{m}\left(\frac{\gagg}{\SI{e-8}{GeV^{-1}}}\right)^{-1}\left(\frac{\mathcal{I}}{\SI{e23}{W/cm^2}}\right)^{-\frac{1}{2}}
\end{equation}
The highest laser intensity currently running systems can generate is around $\mathcal{I}=\SI{e23}{W/cm^2}$ and lasts for $\tau=\SI{10}{fs}$ \cite{Yoon:21}; for a review of current laser technology see \cite{danson2019}. Such intensities would correspond to a wavelength of $\lambda_a=\SI{9}{m}$. Generating and maintaining long laser focuses might be possible with a variety of approaches in the future. For example, recent plasma-waveguides are capable of maintaining a laser focus over $\SI{10}{m}$ scales \cite{2019PhRvS..22d1302S,Picksley:2020qou}. It is also worth pointing out that the focus length decreases mildly with laser power, hence the problem will become marginally simpler in the future. We will for now assume that a sufficiently long focus can be generated to proceed.

The e-folding time is readily found from \eqref{Eq:EfoldGrowth} and must be compared to the pulselength of a $\mathcal{I}=\SI{e23}{W/cm^2}$ pulse which lasts for $\tau\sim\SI{10}{fs}$. This restricts the couplings we can probe, assuming we are capable of measuring a polarisation change $\vartheta$, by inverting \eqref{Eq:PolAngle} we find
\begin{widetext}
\begin{equation}
    \label{Eq:FullSignal}
    \gagg\geq \SI{3.4e-2}{GeV^{-1}}\left(\frac{\mathcal{I}}{\SI{e23}{W/cm^2}}\right)^{-\frac{1}{2}}\left(\frac{\tau}{\SI{10}{fs}}\right)^{-1}\ln\left(\frac{\vartheta}{\varepsilon}\right).
\end{equation}
\end{widetext}
The $\log$ gives an $\mathcal{O}(1)$ contribution for such large values of the coupling; it is not inconceivable to measure polarisation changes $\vartheta <0.01$ therefore $\ln\sim 5$ here. We also note that the axion wavelength for such couplings decreases to $\SI{3}{\mu m}$, a length scale which is very close to the $\SI{1.1}{\mu m}$ spot-size of \cite{Yoon:21}. We therefore conclude that such couplings could be measured today.

\begin{figure}
    \centering
    \includegraphics[width=0.475\textwidth]{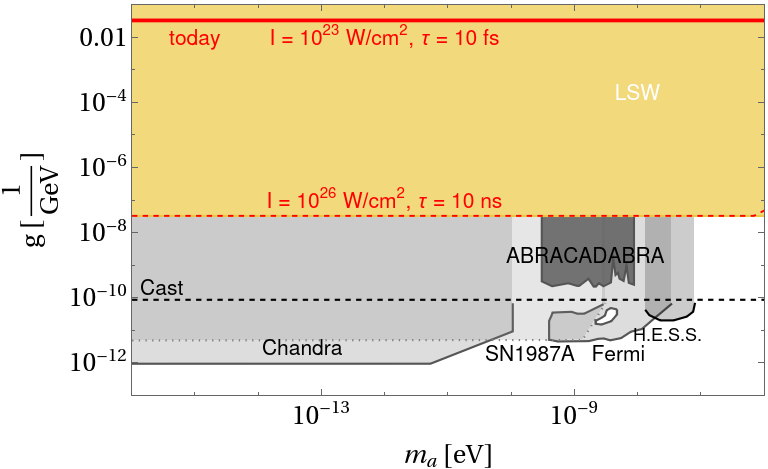}
    \caption{Axion exclusion plot indicating past experiments and the current work. The coloured regions correspond to purely laboratory based experiments while in grey-scale we indicate astrophysical and dark matter bounds. In yellow results from light-shining through wall experiments are shown \cite{Ehret:2010mh,Ballou:2015cka}. The dark grey region is excluded by the ABRACADABRA collaboration \cite{Ouellet:2018beu} looking for dark matter axions. The dashed black line indicates CAST constraints \cite{anastassopoulos2017new}. The lighter grey regions are excluded by the Chandra telescope \cite{Reynolds:2019uqt}, observations made on the SN1987A supernova \cite{Payez:2014xsa}, the Fermi LAT collaboration \cite{TheFermi-LAT:2016zue} and considerations of a Hot Neutron Star in HESS J1731-347 \cite{Beznogov:2018fda}. The solid red line indicates the reach of an experiment as described in the main text with laser intensities of $\mathcal{I}=\SI{e23}{W/cm^2}$ with pulselength $\tau=\SI{10}{fs}$. The dashed red line indicates the laser requirements to reach bounds similar to current LSW bounds.}
    \label{fig:ExclusionPlot}
\end{figure}

The instability cuts off at large masses, hence we are only sensitive to masses satisfying
\begin{equation}
    \label{Eq:MassCutOff}
    m_a\leq\SI{2e-8}{eV}\left(\frac{\gagg}{\SI{e-8}{GeV^{-1}}}\right)\left(\frac{\mathcal{I}}{\SI{e23}{W/cm^2}}\right)^\frac{1}{2}.
\end{equation}
We find that such an approach would reach bounds as indicated by the red line in fig. \ref{fig:ExclusionPlot}. While the bounds are within the exclusion region of LSW experiments it is worth stressing that our bounds on $\gagg$ grow with laser energy and pulse length. To reach current LSW exclusion bounds, we would need to extend the duration of the high intensity pulse to $\SI{10}{ns}$ (without increasing the intensity). Such long pulses are possible, however not currently in combination with high intensities.

It is worth mentioning that this approach can, unfortunately, not be used to probe the QCD axion. Here the coupling and mass are no longer independent but rather $\gagg\propto m_a$. To reach this we would need laser fields above the Schwinger critical field limit.

Another possible application of the axion-photon instability described in this paper is within astrophysics. The very long wavelength of the unstable mode naturally lends itself to macroscopic distances as are usually encountered in astrophysics. Possible applications remain to be investigated.

We would also like to point out that working beyond the perfect phase-matching condition \eqref{Eq:ModeMatching} may lead to an enhanced signal. A phase mismatch may allow shorter wavelengths to become unstable, thus providing a measurable signal in laboratory experiments, at the expense of a lower growth rate. This possibility is left for future work.

\begin{acknowledgments}
The research leading to these results has received funding from AWE plc. British Crown Copyright 2021/AWE.
\end{acknowledgments}

\bibliography{Bibliography}

\begin{thebibliography}{10}

\bibitem{Baker:2006ts}
C.~A. Baker et~al.
\newblock {An Improved experimental limit on the electric dipole moment of the
  neutron}.
\newblock {\em Phys. Rev. Lett.}, 97:131801, 2006.

\bibitem{Afach:2015sja}
J.~M. Pendlebury et~al.
\newblock {Revised experimental upper limit on the electric dipole moment of
  the neutron}.
\newblock {\em Phys. Rev.}, D92(9):092003, 2015.

\bibitem{PhysRevLett.38.1440}
R.~D. Peccei and Helen~R. Quinn.
\newblock $\mathrm{CP}$ conservation in the presence of pseudoparticles.
\newblock {\em Phys. Rev. Lett.}, 38:1440--1443, Jun 1977.

\bibitem{PhysRevD.16.1791}
R.~D. Peccei and Helen~R. Quinn.
\newblock Constraints imposed by $\mathrm{CP}$ conservation in the presence of
  pseudoparticles.
\newblock {\em Phys. Rev. D}, 16:1791--1797, Sep 1977.

\bibitem{Weinberg:1977ma}
Steven Weinberg.
\newblock {A New Light Boson?}
\newblock {\em Phys. Rev. Lett.}, 40:223--226, 1978.

\bibitem{Wilczek:1977pj}
Frank Wilczek.
\newblock {Problem of Strong $P$ and $T$ Invariance in the Presence of
  Instantons}.
\newblock {\em Phys. Rev. Lett.}, 40:279--282, 1978.

\bibitem{Preskill:1982cy}
John Preskill, Mark~B. Wise, and Frank Wilczek.
\newblock {Cosmology of the Invisible Axion}.
\newblock {\em Phys. Lett.}, 120B:127--132, 1983.

\bibitem{Abbott:1982af}
L.~F. Abbott and P.~Sikivie.
\newblock {A Cosmological Bound on the Invisible Axion}.
\newblock {\em Phys. Lett.}, 120B:133--136, 1983.

\bibitem{Dine:1982ah}
Michael Dine and Willy Fischler.
\newblock {The Not So Harmless Axion}.
\newblock {\em Phys. Lett.}, 120B:137--141, 1983.

\bibitem{Witten:1984dg}
Edward Witten.
\newblock {Some Properties of O(32) Superstrings}.
\newblock {\em Phys. Lett.}, 149B:351--356, 1984.

\bibitem{Arvanitaki:2009fg}
Asimina Arvanitaki, Savas Dimopoulos, Sergei Dubovsky, Nemanja Kaloper, and
  John March-Russell.
\newblock {String Axiverse}.
\newblock {\em Phys. Rev.}, D81:123530, 2010.

\bibitem{Wilczek:1987mv}
Frank Wilczek.
\newblock {Two Applications of Axion Electrodynamics}.
\newblock {\em Phys. Rev. Lett.}, 58:1799, 1987.

\bibitem{Sikivie:2020zpn}
Pierre Sikivie.
\newblock {Invisible Axion Search Methods}.
\newblock 3 2020.

\bibitem{Sikivie:1983ip}
P.~Sikivie.
\newblock {Experimental Tests of the Invisible Axion}.
\newblock {\em Phys. Rev. Lett.}, 51:1415--1417, 1983.
\newblock [Erratum: Phys.Rev.Lett. 52, 695 (1984)].

\bibitem{Raffelt:1987im}
Georg Raffelt and Leo Stodolsky.
\newblock {Mixing of the Photon with Low Mass Particles}.
\newblock {\em Phys. Rev.}, D37:1237, 1988.

\bibitem{1986PhLB..175..359M}
L.~{Maiani}, R.~{Petronzio}, and E.~{Zavattini}.
\newblock {Effects of nearly massless, spin-zero particles on light propagation
  in a magnetic field}.
\newblock {\em Physics Letters B}, 175(3):359--363, August 1986.

\bibitem{DellaValle:2015xxa}
Federico Della~Valle, Aldo Ejlli, Ugo Gastaldi, Giuseppe Messineo, Edoardo
  Milotti, Ruggero Pengo, Giuseppe Ruoso, and Guido Zavattini.
\newblock {The PVLAS experiment: measuring vacuum magnetic birefringence and
  dichroism with a birefringent Fabry–Perot cavity}.
\newblock {\em Eur. Phys. J.}, C76(1):24, 2016.

\bibitem{Mendonca:2007zz}
J.~T. Mendonca.
\newblock {Axion excitation by intense laser fields}.
\newblock {\em EPL}, 79(2):21001, 2007.

\bibitem{1988PhRvD..37.1356R}
Georg~G. {Raffelt}.
\newblock {Plasmon decay into low-mass bosons in stars}.
\newblock {\em Phys. Rev. D}, 37(6):1356--1359, March 1988.

\bibitem{Arza:2020zop}
Ariel Arza.
\newblock {Production of massive bosons from the decay of a massless particle
  beam}.
\newblock 9 2020.

\bibitem{Yoon:21}
Jin~Woo Yoon, Yeong~Gyu Kim, Il~Woo Choi, Jae~Hee Sung, Hwang~Woon Lee,
  Seong~Ku Lee, and Chang~Hee Nam.
\newblock Realization of laser intensity over $10^{23}$w$/$cm$^2$.
\newblock {\em Optica}, 8(5):630--635, May 2021.

\bibitem{danson2019}
Colin~N. Danson, Constantin Haefner, Jake Bromage, Thomas Butcher,
  Jean-Christophe~F. Chanteloup, Enam~A. Chowdhury, Almantas Galvanauskas,
  Leonida~A. Gizzi, Joachim Hein, David~I. Hillier, and et~al.
\newblock {Petawatt and exawatt class lasers worldwide}.
\newblock {\em High Power Laser Science and Engineering}, 7:54, 2019.

\bibitem{2019PhRvS..22d1302S}
R.~J. {Shalloo}, C.~{Arran}, A.~{Picksley}, A.~{von Boetticher}, L.~{Corner},
  J.~{Holloway}, G.~{Hine}, J.~{Jonnerby}, H.~M. {Milchberg}, C.~{Thornton},
  R.~{Walczak}, and S.~M. {Hooker}.
\newblock {Low-density hydrodynamic optical-field-ionized plasma channels
  generated with an axicon lens}.
\newblock {\em Physical Review Accelerators and Beams}, 22(4):041302, April
  2019.

\bibitem{Picksley:2020qou}
A.~Picksley et~al.
\newblock {Meter-Scale, Conditioned Hydrodynamic Optical-Field-Ionized Plasma
  Channels}.
\newblock {\em Phys. Rev. E}, 102(5):053201, 2020.

\bibitem{Ehret:2010mh}
Klaus Ehret et~al.
\newblock {New ALPS Results on Hidden-Sector Lightweights}.
\newblock {\em Phys. Lett. B}, 689:149--155, 2010.

\bibitem{Ballou:2015cka}
R.~Ballou et~al.
\newblock {New exclusion limits on scalar and pseudoscalar axionlike particles
  from light shining through a wall}.
\newblock {\em Phys. Rev.}, D92(9):092002, 2015.

\bibitem{Ouellet:2018beu}
Jonathan~L. Ouellet et~al.
\newblock {First Results from ABRACADABRA-10 cm: A Search for Sub-$\mu$eV Axion
  Dark Matter}.
\newblock {\em Phys. Rev. Lett.}, 122(12):121802, 2019.

\bibitem{anastassopoulos2017new}
V~Anastassopoulos, S~Aune, K~Barth, A~Belov, H~Br{\"a}uninger, Giovanni
  Cantatore, JM~Carmona, JF~Castel, SA~Cetin, F~Christensen, et~al.
\newblock New cast limit on the axion--photon interaction.
\newblock {\em Nature Physics}, 13(6):584, 2017.

\bibitem{Reynolds:2019uqt}
Christopher~S. Reynolds, M.C.~David Marsh, Helen~R. Russell, Andrew~C. Fabian,
  Robyn Smith, Francesco Tombesi, and Sylvain Veilleux.
\newblock {Astrophysical limits on very light axion-like particles from Chandra
  grating spectroscopy of NGC 1275}.
\newblock 7 2019.

\bibitem{Payez:2014xsa}
Alexandre Payez, Carmelo Evoli, Tobias Fischer, Maurizio Giannotti, Alessandro
  Mirizzi, and Andreas Ringwald.
\newblock {Revisiting the SN1987A gamma-ray limit on ultralight axion-like
  particles}.
\newblock {\em JCAP}, 02:006, 2015.

\bibitem{TheFermi-LAT:2016zue}
M.~Ajello et~al.
\newblock {Search for Spectral Irregularities due to
  Photon\textendash{}Axionlike-Particle Oscillations with the Fermi Large Area
  Telescope}.
\newblock {\em Phys. Rev. Lett.}, 116(16):161101, 2016.

\bibitem{Beznogov:2018fda}
Mikhail~V. Beznogov, Ermal Rrapaj, Dany Page, and Sanjay Reddy.
\newblock {Constraints on Axion-like Particles and Nucleon Pairing in Dense
  Matter from the Hot Neutron Star in HESS J1731-347}.
\newblock {\em Phys. Rev. C}, 98(3):035802, 2018.

\end{thebibliography}
\bibliographystyle{unsrt}

\end{document}